\documentclass[11pt]{pazh}

\usepackage{psfig}

\usepackage[T2A]{fontenc}

\usepackage{latexsym}
\usepackage{amssymb}

\begin{document}

\setcounter{page}{1}

\sloppypar

\title{\bf Pitch angles of distant spiral galaxies}

\author{S.S. Savchenko, V.P. Reshetnikov} 

\institute{St.Petersburg State University, Universitetskii pr. 28, Petrodvoretz, 
198504 Russia
}

%\authorrunning{ }
\titlerunning{Pitch angles}

\abstract{
We have studied the pitch angles of spiral arms for 31 distant galaxies at
$z \sim 0.7$ from three Hubble Deep Fields (HDF-N, HDF-S, HUDF). Using the 
pitch angle -- rotation velocity relation calibrated from nearby galaxies, 
we have estimated the rotation velocities of galaxies from the deep fields.
These estimates have a low accuracy ($\sim$50 km/s), but they allow low-mass 
and giant galaxies to be distinguished. The Tully--Fisher relation constructed 
using our velocity estimates shows satisfactory agreement with the actually 
observed relations for distant galaxies and provides evidence for the luminosity
evolution of spiral galaxies.
}
\titlerunning{Pitch angles}
\maketitle

\section{Introduction}

The pitch angle of spiral arms is a major parameter
of the classical morphological classification of
galaxies (Hubble 1936). The pitch angle is the angle
between the tangents to the spiral arm and to the
circumference centered at the galaxy nucleus drawn
through a given point. Galaxies with tightly wound
spirals and open arms have small and large pitch
angles, respectively. Typically, the pitch angle lies
within the range $\sim 0^\circ - 30^\circ$ (Kennicutt 1981; Ma 2001).

In 1981, Kennicutt published a paper in which the
relation between the pitch angle of a spiral galaxy
and its maximum rotation velocity was studied. It
followed from this paper that galaxies with tightly
wound arms (smaller pitch angles) rotated, on average,
faster than those with more open arms (larger
pitch angles), with this dependence being linear. In
recent years, investigating the pitch angle of spiral
arms has become increasingly popular, because several
unexplained empirical relationships of this quantity
to the parameters of the galaxy rotation curve and
to the mass of the central black hole have been found
(see, e.g., Seigar et al. 2008; Shields et al. 2010).

The goal of this paper is to develop the results from
Kennicutt (1981) and to try to apply the pitch angle --
rotation velocity relation to determine the rotation
velocities of distant spiral galaxies from the Hubble
Deep Fields HUDF, HDF-N, and HDF-S. Since
many of these galaxies are very faint and have small
angular sizes, measuring their rotation velocities by
spectroscopic methods requires instrumentation that
will not be available in the near future. However,
knowledge of these velocities is needed to study the
evolution of galaxies at high $z$. In addition, there are
virtually no data on the pitch angles of spiral arms for
distant spiral galaxies at present.

All numerical values in
the paper are given for the cosmological model with the
Hubble constant 70 km s$^{-1}$ Mpc$^{-1}$, 
$\Omega_m=0.27$, and $\Omega_{\Lambda}=0.73$.

\section{Measurements}

\subsection{Determining the orientation of galaxy disks}

An important preliminary stage of our work to
study the shape of spiral arms is to estimate the
orientation of galaxy disks in space. This orientation
is specified by two parameters: the inclination $i$
of the disk plane to the plane of the sky and the position
angle $PA$ of the major axis. Knowing these
parameters is needed for two reasons. First, because
of the galaxy inclination to the plane of the sky, the
apparent spiral structure is distorted and, therefore,
the galaxy image should be deprojected to the ``face-on''
orientation. Second, the galaxy rotation velocity
derived from spectroscopic measurements should
also be corrected for the inclination.

In this paper, we decided to use the relatively new
method of spiral-arm monotony (SAM; Poltorak and
Fridman 2007; Fridman and Poltorak 2010). This
method is based on the assumption that every spiral
arm is a monotonic function, i.e., following along
the spiral from the galaxy center to the periphery,
the radius must increase monotonically
($\frac{dr}{d\phi}>0$, where $r$ is the distance from a 
point in the spiral arm to the galaxy center and
$\phi$ is the azimuthal angle). In this case, the spiral 
projected onto the plane
of the sky can be represented by a nonmonotonic
function. Thus, the domain of $i$ and $PA$ for which the
deprojected spiral is monotonic will be the domain of
possible inclinations and position angles.

An example of using the SAM method to estimate
the orientation of four nearby galaxies is shown in
Fig.~1. We see from the figure that the domains of
possible values are relatively small, which allows both
inclination and position angle of the galaxy major axis
to be estimated with a good accuracy.

\begin{figure*}
\centerline{\psfig{file=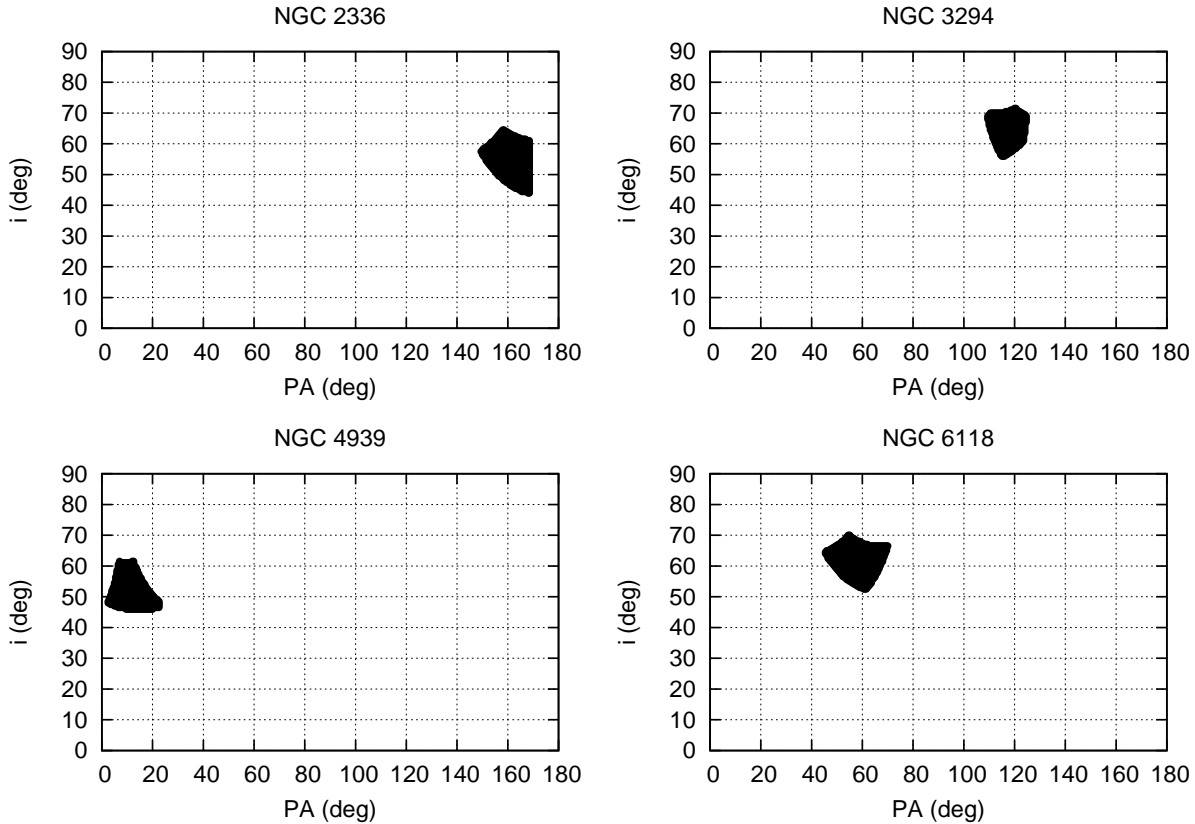,width=16cm,angle=-90,clip=}}
\caption{Results of the operation of the spiral-arm monotony method 
for the galaxies NGC 2336, NGC 3294, NGC 4939, and
NGC 6118 as an example. The black color indicates the domain of 
possible inclinations and position angles.}
\end{figure*} 

\subsection{The pitch angle}

Once the galaxy images have been corrected for
the inclination, we can turn to the determination
of spiral-arm pitch angles. As the published data
show, the pitch angles for the same objects often
differ markedly. Therefore, for reliability, we decided
to implement two different, completely independent
methods and to compare their results.

The first method (below called the interactive one)
is based directly on the search for the angle between
the tangents to the spiral arm and to the circumference
centered at the galaxy nucleus. If the spiral
arm is represented in polar coordinates: $r = r(\phi)$ 
(the coordinate origin at the galaxy center), then the pitch
angle for a logarithmic spiral can be determined from
the formula (see, e.g., Binney and Tremaine 1987)
\begin{equation}
\mu = {\rm arcctg} \left( r \left| \frac{d \phi }{d r} \right| \right).
\end{equation}

The main steps in determining the pitch angle are:
(1) determining the coordinates of the galaxy center;
(2) finding the polar coordinates of several ($\sim$10)
points on the arm; (3) applying pairwise Eq.~(1) to the
derived coordinates to obtain the set of pitch angles
corresponding to different arm segments; (4) averaging
the pitch angles to obtain the final result.

To reduce the random errors, this procedure was
repeated several times for each arm. If the galaxy
had several (as a rule, two) large-scale arms suitable
for measurements, then their pitch angles were also
estimated in the same way. As the final pitch angle
for a given galaxy, we took the value averaged over
the measured arms.

The second method is based on a Fourier analysis
of the distribution of points in the spiral arms of a
galaxy (Considere and Athanassoula 1982). If the
distribution of points in the galaxy arms is represented
as the sum of delta functions of their polar coordinates:
\begin{equation}
\frac{1}{N} \sum_{i=1}^N \delta(u-u_i) \delta(\phi-\phi_i),
\end{equation}
where $u_i=\ln(r_i)$, then its Fourier transform
\begin{equation}
A(p,m) = \int_{-\infty}^{\infty} \int_{-\pi}^{\pi} \frac{1}{N} \sum_{i=1}^N \delta(u-u_i) \delta(\phi-\phi_i)
\end{equation}
$$
\times e^{-i(pu+m\phi)} \, dud\phi 
$$
gives the coefficients in the expansion of this distribution
in terms of logarithmic spirals. The pitch angle
can be found from the formula
\begin{equation}
\mu = \arctan \left( - \frac{m}{p_{max}}\right),
\end{equation}
where $m$ is the number of spiral arms in the galaxy
and $p_{max}$ is the value of the parameter $p$ at which the
function $|A(p,m)|$ has a maximum.

\section{Measuring the pitch angles for galaxies of a local sample}

\subsection{The galaxy sample}

To study the spiral pattern of nearby galaxies,
we used the galaxy sample described by Kennicutt
(1981). The sample includes 113 spiral galaxies
of various types with a clearly distinguishable spiral
structure. Since the arms of barred galaxies are
described more poorly by a logarithmic spiral, most
of the sample galaxies are without bars or have small
bars.

The galaxy images in the fits format were retrieved
from the NED\footnote{NASA/IPAC Extragalactic Database}, 
which contains references to observations
with different instruments and in different
spectral ranges. The pitch angle depends on the filter;
for our purposes, we used only the $B$-band images,
because the spiral structure is seen best in blue bands.
No photometric calibration of the galaxy images was
required and, therefore, we used the data obtained
with different instruments; from several possible options,
we chose the best one from the viewpoint of
visual image reduction (where possible, a higher resolution
and a better view of the spiral structure).

The maximum rotation velocities of the sample
galaxies $V_{max}$ found from the width of the
HI ($\lambda$=21 cm) line profile were retrieved from the 
HyperLEDA\footnote{Lyon-Meudon Extragalactic Database}.
The HyperLEDA rotation velocity have already
been corrected for the inclination found from the apparent
flattening of galaxies and, therefore, we initially
eliminated this correction and subsequently applied it
again using the galaxy inclination estimated by the
SAM method (see below).

\subsection{Measuring the pitch angle}

For all galaxies from the sample by Kennicutt
(1981), we estimated their orientation parameters,
the inclination and position angle of the major axis,
by the SAM method. The agreement of our inclinations
with those in HyperLEDA is, on average,
good:  $\langle i_{MCP} - i_{LEDA} \rangle = -2^\circ \pm 13^\circ$.
However, for several galaxies, the apparent isophote flattening
method gives, obviously, incorrect values due to various
kinds of peculiarities in the shapes of the outer
isophotes of galaxies.

The SAM method has a limitation at low inclinations
(i.e., when the galaxies are seen almost face-on):
the error in the $i$ and $PA$ estimates increases
greatly with decreasing inclination. For this reason,
we failed to estimate the orientation parameters for
16 galaxies and they were excluded from our sample,
in which, thus, 97 objects remained.

Once the images have been deprojected, we determined
the spiral-arm pitch angle for each galaxy by
the two methods. At this step, we excluded several
more galaxies with an excessively irregular and asymmetric
structure from our sample. We failed to find
the maximum rotation velocities for four galaxies. In
addition, we decided to restrict ourselves to the galaxies
with inclinations in the range $30^\circ \leq i \leq 60^\circ$,
because both pitch angles and maximum rotation
velocities can be reliably estimated for such galaxies
and their pitch angle -- rotation velocity relation (see
below) is seen much better. As a result, 46 galaxies
remained in our sample of nearby galaxies with pitch
angles measured by the two methods and with known
rotation velocities, 43 and 3 of which have two-armed
and three-armed spiral patterns, respectively.

\begin{figure*}
\centerline{\psfig{file=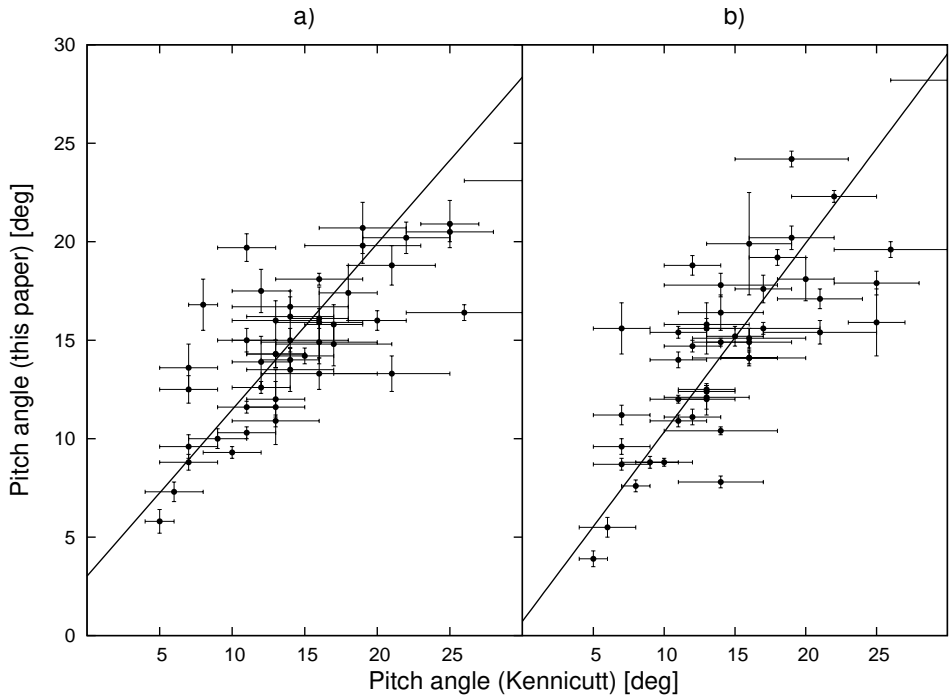,width=13cm,angle=-90,clip=}}
\caption{Comparison of our pitch angles with those from Kennicutt (1981):
(a) for the interactive method; (b) for the method
based on a Fourier analysis of galaxy images.}
\end{figure*} 

Figure 2 compares our pitch-angle measurements
with the results from Kennicutt (1981) for both methods.
The straight lines indicate linear fits to the data.
They suggest the existence of certain systematics in
the pitch-angle measurements probably attributable
to the difference between the methods used. The
mean differences between the measurements for the
same galaxies are small:
$\Delta\mu = 0.2^\circ \pm 3.7^\circ$ (the
difference between the pitch angle measured by the
first (interactive) method and the angle measured by
Kennicutt) and
$\Delta\mu = 0.0^\circ \pm 3.5^\circ$ (the same for
the second method). 

Comparison of the two pitch-angle determination
methods described above shows good mutual agreement
between the results: the difference between
the angles estimated by the Fourier and interactive
methods is $\Delta\mu = -0.2^\circ \pm 3.0^\circ$. 
Both methods yield similar results, although, formally, the accuracy
of the method based on a Fourier analysis is higher.

Figure~3 shows the logarithmic spirals constructed
from the spiral-pattern parameters we 
determined and, given with the galaxy inclinations
and position angle, they were superimposed on the
observed images of NGC~2997 and NGC~4254.

\begin{figure*}
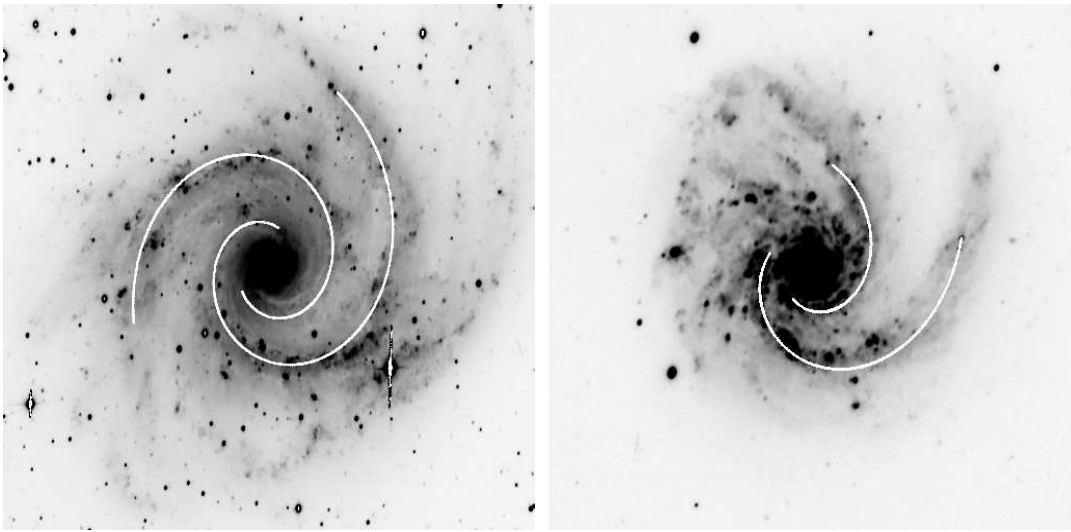

\centering
$    \begin{array}{cc}
\psfig{file=fig3a.eps,width=7.0cm,clip=} &
\psfig{file=fig3b.eps,width=7.0cm,clip=} \\
\end{array} $
\caption{Images of the galaxies NGC~2997 (left) and NGC~4254 (right) 
with the logarithmic spirals corresponding to the pitch angles
measured by the interactive method superimposed on them.}
\end{figure*}
                     
\subsection{The pitch angle -- rotation velocity relation for
nearby galaxies}                       

Figure~4 shows the pitch angle -- rotation velocity
relations constructed by the two methods. As we
see from the figure, both approaches give significant
correlations. However, the relation constructed from
the measurements using the Fourier transform has a
higher statistical significance (the linear correlation
coefficient for it is 0.725 versus 0.600 for the relation
constructed by the interactive method). In our subsequent
discussion, we will use the pitch angles found
by a Fourier analysis. Note also that the galaxies with
a three-armed pattern in Fig.~4 are near the average
relation for galaxies with a two-armed pattern.

\begin{figure*}
\centerline{\psfig{file=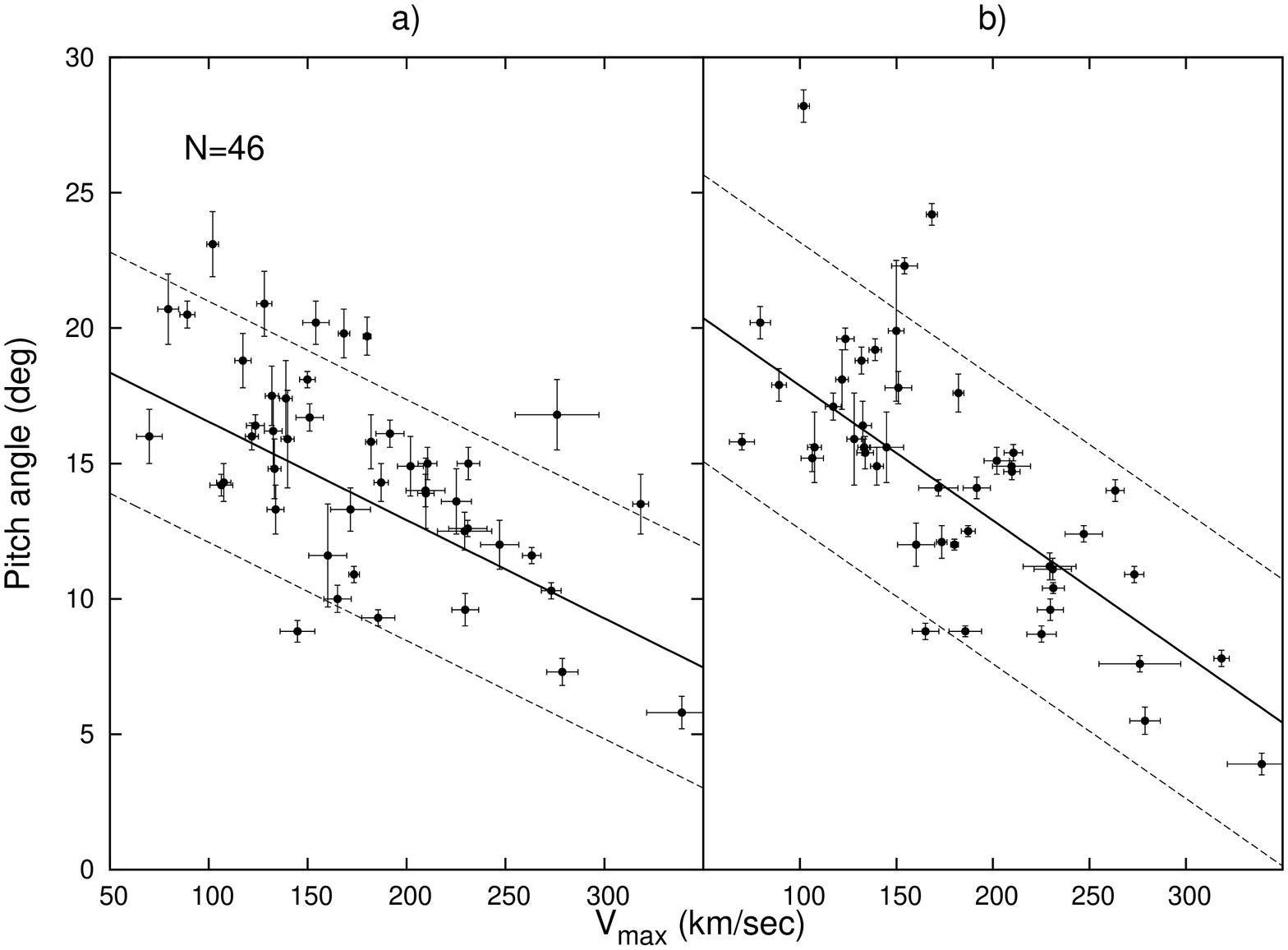,width=13cm,angle=-90,clip=}}
\caption{Pitch angle -- maximum rotation velocity relations derived by 
the two methods: the interactive method (a) and Fourier
analysis (b). The solid lines are linear regressions; the dashed lines are
$\pm\sigma$ deviations.}
\end{figure*} 

The final empirical relation found from 46 nearby
galaxies (the solid straight line in Fig. 4b) is
\begin{equation}
\mu (^\circ) = (-0.049 \pm 0.008) \cdot V_{max} {\rm (km/s)} 
\end{equation}
$$
~~~~~~~~~~~~~~~~~~~~~~~~~~~~+~ (22.85 \pm 1.76).
$$
    
Curiously, our Galaxy may satisfy this relation. It
follows from Eq.~(5) that $\mu = 12^\circ$ for the Milky Way
with $V_{max}=220$ km/s. This pitch angle is close
to the estimates of various authors (see, e.g. Tables
1 and 2 in Vallee 2005). However, greatly differing
pitch angles of the Milky Way spiral pattern are
also encountered in the literature (see, e.g., Levine
et al. 2006).

\section{Measuring the pitch angles for galaxies from the deep fields}

\subsection{The galaxy sample}

To find distant galaxies with a spiral structure, we
used the original frames of the Hubble Deep Fields
HDF-N and HDF-S (Ferguson et al. 2000) as well
as the Ultra Deep Field HUDF (Beckwith et al. 2006).
The main problem in compiling the sample was that
the angular sizes of many distant galaxies are too
small and, therefore, several objects with a distinct
spiral structure were not included in the sample to
avoid the errors due to image discreteness. In addition,
a considerable number of deep-field galaxies
appear asymmetric and peculiar and no regular spiral
arms can be drawn for them.

Our final sample includes 31 galaxies (20 in
HUDF, six in HDF-N, and five in HDF-S) with a
clearly distinguishable spiral pattern, as a rule, a two-armed
ones as in the case of nearby galaxies. The
mean redshift of these galaxies is 
$\langle z \rangle = 0.69\pm0.30$.
Our list of distant galaxies is presented in the table.

\subsection{Measurements}

For all sample galaxies, we downloaded their images
in the F606W filter, because this is the only
common filter for all three fields. To estimate the
orientation of distant galaxies, we used the SAM
method. In the cases where this method yielded
no definite results, we estimated the inclination and
orientation from the outer isophotes of the galaxies.
(Such objects were discarded among the nearby
galaxies but were retained among the distant galaxies,
because of the small size of their sample. Their
inclinations are marked by the asterisk in the table.)

The results of our measurements of the inclination ($i$)
and spiral-pattern pitch angle ($\mu$) are summarized in
the table. If a galaxy has only one arm suitable for
measurements, the results are presented only for it.
The galaxy redshifts (the fourth column of the table)
were taken from Wolf et al. (2004) (the COMBO-17
project) for HUDF, from Wirth et al. (2004) for HDF-N,
and from Sawicki and Mallen-Ornelas (2003) for
HDF-S. (We used the redshift for the HUDF galaxy
No. 16 from Coe et al. (2006), because $z$ from Wolf
et al. (2004) leads to an unrealistically low luminosity
of the galaxy.) The eighth column of the table
presents the maximum rotation velocity found from
the empirical relation (5) and its error. We estimated
the rotation-velocity error using the formula for the
propagation of the mean error (the formula for the
error of a function of several variables) from the pitch angle
measurement error and the errors of the numerical
coefficients in Eq.~(5). The fifth column gives
the absolute magnitudes of the galaxies in the Hubble
Space Telescope F606W filter. To find the absolute
magnitudes, we used the $k$-corrections for galaxies of
the corresponding types from Bicker et al. (2003).

%   Table

\begin{table*}
\caption{Galaxies from the deep fields}
\begin{center}
\begin{tabular}{|c|c|c|c|c|c|c|c|}
\hline
No. & RA          &     DEC      & $z$    &$M_{F606W}$& $i$ &   $\mu$       &   $V_{max}$  \\ 
  &             &              &      &           &($^\circ$)& ($^\circ$)   &   (km/s)     \\
\hline
\multicolumn{8}{|c|}{HUDF} \\ 
\hline
 1 &03:32:39.26 & -27:45:32.37 & 0.917 & -20.76 & 37   & 16.4 $\pm$ 0.4 & 129 $\pm$ 42 \\
 2 &03:32:42.81 & -27:46:05.72 & 0.667 & -22.35 & 58   & 17.8 $\pm$ 0.4 & 101 $\pm$ 40 \\
 3 &03:32:35.77 & -27:46:27.62 & 1.007 & -20.52 & 57   &  8.9 $\pm$ 0.5 & 280 $\pm$ 60 \\
 4 &03:32:38.97 & -27:46:30.30 & 0.457 & -21.05 & 49   & 12.5 $\pm$ 0.9 & 208 $\pm$ 53 \\
 5 &03:32:46.10 & -27:47:13.94 & 1.122 & -21.97 & 46   & 17.8 $\pm$ 0.7 & 101 $\pm$ 41 \\
 6 &03:32:39.87 & -27:47:14.98 & 1.086 & -23.00 & 41   & 15.5 $\pm$ 0.7 & 147 $\pm$ 45 \\
 7 &03:32:34.10 & -27:47:12.13 & 0.153 & -18.24 & 76   & 20.3 $\pm$ 1.3 &  51 $\pm$ 37 \\
 8 &03:32:31.35 & -27:47:24.92 & 0.656 & -17.90 & 68   & 19.2 $\pm$ 0.7 &  93 $\pm$ 44 \\
 9 &03:32:44.86 & -27:47:27.65 & 0.187 & -20.75 & 58   & 12.0 $\pm$ 0.7 & 218 $\pm$ 53 \\
10 &03:32:45.07 & -27:47:38.65 & 0.349 & -20.76 & 70   & 15.8 $\pm$ 1.6 & 141 $\pm$ 53 \\
11 &03:32:39.17 & -27:48:44.63 & 0.472 & -22.36 & 56   &  9.3 $\pm$ 0.8 & 272 $\pm$ 61 \\
12 &03:32:34.52 & -27:48:48.38 & 0.236 & -21.40 & 43   & 11.5 $\pm$ 1.0 & 228 $\pm$ 56 \\
13 &03:32:42.28 & -27:47:46.16 & 0.939 & -22.65 & 48   & 16.3 $\pm$ 0.8 & 131 $\pm$ 45 \\
14 &03:32:43.25 & -27:47:56.18 & 0.677 & -21.95 & 46   & 15.2 $\pm$ 0.4 & 153 $\pm$ 44 \\
15 &03:32:41.34 & -27:45:54.42 & 0.533 & -20.42 &  17* & 19.5 $\pm$ 0.7 &  67 $\pm$ 39 \\
16 &03:32:38.34 & -27:45:44.29 & 1.314 & -21.46 &  29* & 10.2 $\pm$ 0.7 & 254 $\pm$ 58 \\
17 &03:32:40.78 & -27:46:15.72 & 0.627 & -22.41 &  33* & 14.6 $\pm$ 0.8 & 165 $\pm$ 48 \\
18 &03:32:33.04 & -27:47:30.89 & 1.064 & -22.25 &  33* & 15.4 $\pm$ 1.9 & 162 $\pm$ 49 \\
19 &03:32:37.87 & -27:47:51.13 & 0.795 & -21.16 &  25* & 10.2 $\pm$ 1.3 & 254 $\pm$ 62 \\
20 &03:32:39.80 & -27:46:53.57 & 0.996 & -21.23 &  26* & 17.9 $\pm$ 0.9 &  99 $\pm$ 43 \\
\hline
\multicolumn{8}{|c|}{HDF-S} \\ 
\hline
 1 &22:32:47.57 & -60:34:08.59 & 0.579 & -21.41 & 52   & 11.1 $\pm$ 0.8 & 236 $\pm$ 46 \\
 2 &22:33:03.57 & -60:33:41.67 & 0.734 & -23.50 & 45   & 10.6 $\pm$ 1.1 & 249 $\pm$ 56 \\
 3 &22:32:47.65 & -60:33:35.87 & 0.581 & -22.92 & 50   &  8.1 $\pm$ 1.1 & 296 $\pm$ 66 \\
 4 &22:33:00.24 & -60:32:34.03 & 0.415 & -21.04 & 41   & 14.2 $\pm$ 0.8 & 173 $\pm$ 49 \\
 5 &22:32:57.99 & -60:32:34.32 & 0.761 & -21.83 & 45   & 16.0 $\pm$ 1.4 & 137 $\pm$ 51 \\
\hline
\multicolumn{8}{|c|}{HDF-N} \\ 
\hline
 1 &12:36:45.86 & +62:13:25.87 & 0.320 & -19.91 & 29   &  7.9 $\pm$ 1.4 & 300 $\pm$ 68 \\
 2 &12:36:48.72 & +62:13:19.40 & 0.753 & -20.73 & 16   & 19.1 $\pm$ 1.1 &  75 $\pm$ 43 \\
 3 &12:36:50.22 & +62:12:39.74 & 0.474 & -21.42 & 57   & 13.6 $\pm$ 1.0 & 185 $\pm$ 51 \\
 4 &12:36:56.65 & +62:12:45.32 & 0.518 & -21.59 & 67   & 14.0 $\pm$ 1.1 & 177 $\pm$ 51 \\
 5 &12:36:46.14 & +62:11:43.10 & 1.016 & -23.07 &  0*  & 13.6 $\pm$ 0.6 & 185 $\pm$ 49 \\
 6 &12:36:43.18 & +62:11:48.00 & 1.007 & -21.45 & 35   & 12.0 $\pm$ 0.4 & 218 $\pm$ 52 \\
\hline
\end{tabular}
\end{center}
\end{table*}

\section{Comparison of the local and distant galaxies}

Figure~5 presents the histograms showing the
pitch-angle distributions for the local-sample galaxies
and for the galaxies from the deep fields.
We see good agreement in the distribution
of angles: a maximum near 14$^\circ$ and a gradual decline
on both sides of the maximum. Thus, one might expect
the distributions of rotation velocities for galaxies
at different $z$ to be similar.

\begin{figure}
\centerline{\psfig{file=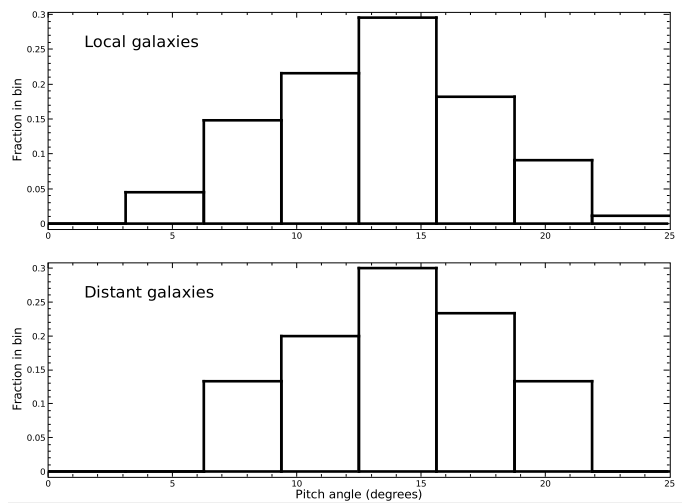,width=9cm,angle=-90,clip=}}
\caption{Histograms representing the pitch-angle distributions of galaxies: 
for nearby galaxies (top) and for distant galaxies from the
deep fields (bottom). The fractions of galaxies in the corresponding bins are shown 
along the vertical axes.}
\end{figure} 

Using the maximum rotation velocities of distant
galaxies estimated from the empirical relation (5) (the
table), we constructed their Tully--Fisher (TF) relation
(Fig.~6). The solid line in the figure indicates the
relation for nearby galaxies in the same color band
(Sakai et al. 2000). As we see from the figure, the
distant galaxies are located on this plane with approximately
the same slope as that for the nearby galaxies,
but their distribution appears shifted towards higher
luminosities. The value of this shift depends on the
galaxy redshift: the galaxies at $z<0.66$, on average,
follow the local relation, while those at $z>0.66$
are shifted upward (Fig.~6).

\begin{figure*}
\centerline{\psfig{file=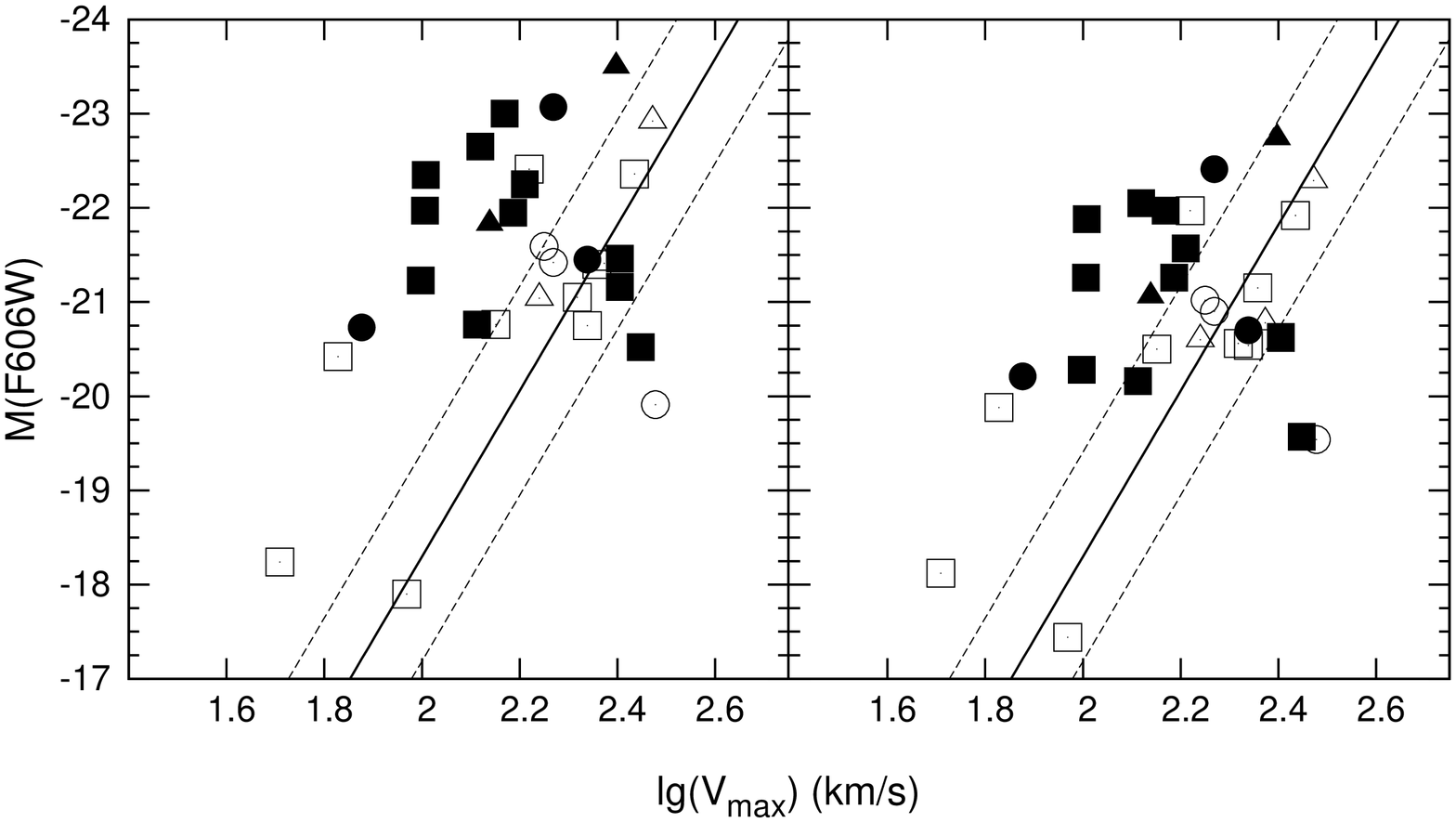,width=15cm,angle=-90,clip=}}
\caption{The Tully--Fisher relation for distant galaxies derived using 
the pitch angle -- rotation velocity relation: the absolute
magnitudes were calculated by applying only the $k$-correction (left) 
and the correction for evolution (right). The straight lines
correspond to the local relation (Sakai et al. 2000); the dashed 
straight lines indicate the $\pm 3\sigma$ deviations. The circles, triangles,
and squares represent the galaxies from HDF-N, HDF-S, and HUDF, respectively. 
The open and filled symbols indicate the galaxies with $z < 0.66$
and $z > 0.66$, respectively.} 
\end{figure*}

The most probable cause of the systematic shift
in the positions of distant galaxies is the evolution of
their luminosity. As was shown in numerous works
devoted to studying the evolution of the TF relation
and the galaxy luminosity function, spiral galaxies at
$z \sim 1$ are brighter than nearby galaxies with the same
maximum rotation velocity by $\sim 1^m$ (see, e.g., Table~1
in Portinari and Sommer-Larsen 2007). The right
panel of Fig.~6 shows the TF relation corrected for the
luminosity evolution of distant galaxies (the evolution
correction was applied as prescribed by Bicker et
al. 2003). As we see from the figure, allowance for
the luminosity evolution slightly improved the agreement
between the relations for distant and nearby
galaxies.

Consider two subsamples approximately equal
in size -- galaxies with $z < 0.66$ (15 objects) and
$z > 0.66$ (16 galaxies). The mean characteristics
of the first subsample are  $\langle z \rangle = 0.44 \pm 0.0
4$ (the standard error of the mean is given),
$\langle M_{F606W} \rangle = -20.91 \pm 0.36$
(only the $k$-correction was applied), and
$\langle V_{max} \rangle = 187 \pm 20$ km/s; 
the characteristics of the second subsample are
$\langle z \rangle = 0.93 \pm 0.05$, 
$\langle M_{F606W} \rangle = -21.87 \pm 0.22$,
and $\langle V_{max} \rangle = 167 \pm 16$ km/s.

Using the local TF relation from Sakai
et al. (2000), we can estimate the expected absolute
magnitude for a galaxy with $V_{max} = 187 \pm 20$ km/s
to be $M_{F606W} = -20.7 \pm 1.0$ and for a galaxy with
$V_{max} = 167 \pm 16$ km/s to be $M_{F606W} = -20.3 \pm 0.9$.
The mean observed magnitude for galaxies with
$\langle z \rangle = 0.44$ is brighter than the expected one by
$\Delta M = 0.^m2 \pm 1.^m1$ and for the more distant subsample
by $\Delta M = 1.^m6 \pm 0.^m9$. According to the model
by Bicker et al. (2003), the corresponding values
of the luminosity evolution for Sb-Sc galaxies are
0.$^m$3--0.$^m$5 and 0.$^m$6--0.$^m$9. Given the approximate
nature of our method for estimating the rotation
velocities and the small size of the galaxy sample, it
may be concluded that the shift of distant galaxies in
the TF relation observed in Fig.~6 agrees satisfactorily
with that expected from the luminosity evolution
with $z$.

\section{Conclusions}

We determined the spiral-arm pitch angles for
46 nearby galaxies with $30^\circ \leq i \leq 60^\circ$
by two different methods (the interactive one and using a Fourier
analysis of images). The inclination of the galaxy
plane to the line of sight was found by the relatively
new method of spiral-arm monotony (Poltorak and
Fridman 2007; Fridman and Poltorak 2010).

We confirmed the conclusion by Kennicutt (1981)
about the existence of a significant correlation between
the rotation velocity of a galaxy and the pitch
angle of its spiral arms. This correlation is best traced
for intermediate galaxy disk inclinations, when both
pitch angles and rotation velocities can be found with
a good accuracy.

We measured the pitch angles for 31 spiral galaxies
from several Hubble Deep Fields at mean redshift
$\langle z \rangle \approx 0.7$. 
Using the local empirical pitch angle -- rotation
velocity relation, we estimated the maximum rotation
velocities of distant galaxies (the table).
These estimates have a low accuracy ($\approx50$ km/s,
see the table), but they allow low-mass and giant
galaxies to be confidently distinguished. 

We constructed the TF relation between the absolute
magnitudes and maximum rotation velocities of
distant galaxies estimated from the pitch angles of
the spiral pattern. Despite the large scatter of data,
we can tentatively conclude that the distant galaxies
follow the local TF relation with approximately the
same slope and, in addition, show evidence of luminosity
evolution. These results agree with the direct
measurements of the TF relation for distant galaxies
by spectroscopic methods.

The ``morphological'' estimates of the rotation velocities
for distant galaxies can be useful in studying
the evolution of spiral galaxies, especially in the
cases where no data on the rotation of a galaxy can
be obtained by spectroscopic methods because of its
faintness or its visibility conditions (the galaxy is seen
almost face-on).

\section*{REFERENCES}

\indent

1. S.V.W. Beckwith, M. Stiavelli, A.M. Koekemoer, et
al., Astron. J. 132, 1729 (2006).

2. J. Bicker, U. Fritze-v. Alvensleben, C.S. Moller, and
K. J. Fricke, Astron. Astrophys. 413, 37 (2004).

3. J. Binney and S. Tremaine, Galactic Dynamics
(Princeton Univ. Press, Princeton, 1987).

4. D. Coe, N. Benitez, S.F. Sanchez, et al., Astron.
J. 132, 926 (2006).

5. S. Considere and E. Athanassoula, Astron. Astrophys.
111, 28 (1982).

6. H.C. Ferguson, M.Dickinson, and R. Williams, Ann.
Rev. Astron. Astrophys. 38, 667 (2000).

7. A.M. Fridman and S.G. Poltorak, Mon. Not. R. Astron.
Soc. 403, 1625 (2010).

8. E. Hubble, The Realm of the Nebulae (Oxford Univ. Press, 1936).

9. R.C. Kennicutt, Jr., Astron. J. 86, 1847 (1981).

10. E.S. Levine, L. Blitz, and C. Heiles, Science 312,
1773 (2006).

11. J. Ma, Chin. J. Astron. Astrophys. 1, 395 (2001).

12. S.G. Poltorak and A.M. Fridman, Astron. Rep. 51, 460 (2007).

13. L. Portinari and J. Sommer-Larsen, Mon. Not.
R. Astron. Soc. 375, 913 (2007).

14. S. Sakai, J.R. Mould, Sh.M.G. Hughes et al., Astrophys.
J. 529, 698 (2000).

15. M. Sawicki and G. Mallen-Ornelas, Astron. J. 126,
1208 (2003).

16. M.S. Seigar and P.A. James, Mon. Not. R. Astron.
Soc. 299, 685 (1998).

17. M.S. Seigar, D. Kennefick, J. Kennefick, and
C.H.S. Lacy, Astrophys. J. 678, L93 (2008).

18. D.W. Shields, J.A. Hughes, R.S. Barrows, et al.,
Bull. Am. Astron. Soc. 42, 381 (2010).

19. J.P. Vallee, Astron. J. 130, 569 (2005).

20. G.D. Wirth, Ch.N.A. Willmer, and P. Amico, Astron.
J. 127, 3121 (2004).

21. C. Wolf, K. Meisenheimer, A. Borch, et al., Astron.
Astrophys. 421, 913 (2004).

\end{document}